\documentclass[%
 reprint,
 amsmath,amssymb,
 aps,
]{revtex4-2}

\usepackage{graphicx}
\usepackage{dcolumn}
\usepackage{bm}
\usepackage{color}

\begin{document}

\preprint{APS/123-QED}

\title{Quantum Kinetic Equation for the Wigner Function \\and Reduction to the Boltzmann Transport Equation under Discrete Impurities
}

\author{Nobuyuki Sano}
\affiliation{%
 Institute of Applied Physics, University of Tsukuba \\
 1-1-1 Tennoudai, Tsukuba, Ibaraki 305-8573, Japan 
}%




\date{\today}

\begin{abstract}
We derive a quantum kinetic equation under discrete impurities for the Wigner function from the quantum Liouville equation. 
To attain this goal, the electrostatic Coulomb potential is separated into the long- and short-range parts, and the self-consistent coupling with Poisson's equation is explicitly taken into account.
It is shown that the collision integral associated with impurity scattering as well as the usual drift term is derived on an equal footing and that the conventional treatment of impurity scattering under the Wigner function scheme is inconsistent in the sense that the collision integral is introduced in an {\it ad hoc} way and, thus, the short-range part of the impurity potential is double-counted. The Boltzmann transport equation (BTE) is derived without imposing an assumption of random impurity configurations over the substrate. Therefore, the derived BTE is able to describe the discrete nature of impurities such as potential fluctuations and, thus, appropriate to analyze electron transport under semiconductor nanostructures.
\end{abstract}

\maketitle


\section{\label{sec:intro}Introduction}
Electron transport properties associated with impurity scattering in doped semiconductors are generally studied by the semiclassical BTE coupled self-consistently to Poisson's equation~\cite{Jacoboni1983,Vasileska2006,Jacoboni2010book,Fischetti2016book}. 
However, such theoretical approaches based on the BTE are incomplete in some respects, especially when taking into account the discrete nature of impurities.
For instance, electrostatic Hartree potential determined by Poisson's equation represents the potential under the long-wavelength limit~\cite{Sano2018} and impurity scattering in the collision integral is evaluated under the implicit assumption of self-averaging over all possible impurity configurations~\cite{Jacoboni2010book,Fischetti2016book,Sano2018,Chattopadhyay1981,SANO2017sse,Pronin2019}. 
We would also like to mention that the collision integral in the BTE is introduced by the reasoning that the rate of change of the electron distribution function must be balanced between drift and scattering processes~\cite{Kohn1957,Greenwood1958,Huang1987}. 
As a result, the conventional theoretical framework based on the BTE completely ignores the discreteness of impurities and cannot describe the phenomena associated with it such as potential fluctuations.  

A quantum mechanical approach based on the Wigner function does not resolve the problem described above; the classical electron distribution function is replaced by the Wigner function and the Hartree potential is replaced by the quantum potential derived from the quantum Liouville equation under the long-wavelength limit~\cite{Jacoboni2010book,Ferry2018}. The scattering processes are introduced through the collision integral with the same reasoning as that in the BTE~\cite{Frensley1987a,Nedjalkov2004,Querlioz2006,Sverdlov2005}. 
We should also mention that the conventional framework of the Wigner function is inconsistent from both mathematical and physical perspectives: The Wigner function cannot be interpreted as a probability density because it is not positive semi-definite, and the scattering potential of impurities is double-counted in both the drift term through the Hartree potential and the collision integral, as we shall show in the present paper.

Therefore, under the conventional framework based on the BTE and the Wigner function scheme, impurity scattering is treated independently from the transport equation. The transition probability due to impurity scattering is evaluated by Fermi's golden rule with the {\it a priori} screened Coulomb potential. 
This implicitly implies that impurities are always fully screened by carriers no matter whether the system is in equilibrium or nonequilibrium.
In other words, as far as impurity scattering is concerned, the ``self-consistent coupling'' between the transport equation and Poisson's equation is not actually consistently coupled under the present framework. 

In fact, the discrete nature of impurities plays a crucial role in the state-of-the-art technology associated with electron devices: Because of drastic miniaturization of device size, electron transport is now quasi-ballistic~\cite{Sano2004b} and the discrete nature of doped impurities is manifested in various transport characteristics~\cite{Sano2017SISPAD}. Especially,
discrete impurities induce local potential fluctuations around impurities. Such potential fluctuations are usually masked by screening carriers and do not show up explicitly in equilibrium under bulk structures. This is why the long-wavelength limit imposed on the electrostatic potential in Poisson's equation is justified under the conventional framework~\cite{Sano2017SISPAD}. 
However, electron devices operate under nonequilibrium conditions and the semiconductor substrate is strongly inhomogeneous. As a result, screening by free carriers is in most cases incomplete and potential fluctuations always show up during device operation. This leads to threshold voltage variabilities (called random dopant fluctuations), which cannot be artificially controlled due to the random nature of impurity configurations. Such variabilities in device characteristics are considered to be a dominant factor preventing further miniaturization of the leading-edge electron devices~\cite{Hoeneisen1972,Keyes1975,Nishinohara1992,Wong1993,Sano2000,Sano2002}.  

In the present paper, we discuss the quantum and semiclassical transport equations with impurity scattering from two different perspectives: One is about the inconsistency between the kinetic equation for the Wigner function and Poisson's equation that exists in double-counting the impurity scattering potential. The other is about the implementation of the discrete nature of impurities into the BTE framework.
These issues may seem to be uncorrelated, but actually, they are intimately linked. The relationship between the self-consistent Hartree potential in the drift term and the scattering potential in the collision integral becomes clearer only if the discrete nature of impurities is explicitly introduced into the kinetic equation. 
To achieve this goal, we separate the long- and short-range parts of the impurity potential in both the quantum kinetic equation for the Wigner function and Poisson's equation and the self-consistent coupling between the two equations are explicitly taken into account. 
It is shown that the collision integral and the self-consistent Hartree potential can be naturally {\em derived} in the quantum kinetic equation on an equal footing. 
Then, the semiclassical BTE is reduced from the quantum kinetic equation without relying on self-averaging of random impurity configurations, which is a key ingredient to derive the BTE for impurity scattering under the long-wavelength limit~\cite{Kohn1957,Greenwood1958,Jancel2013book,Kreuzer1981book}. 

The present paper is organized as follows. 
In Sec.~\ref{sec:Wigner}, the Wigner function is defined and explicit separation of impurity potential into the long- and short-range parts is introduced in the theoretical framework. In Sec.~\ref{sec:Kinetic}, the quantum kinetic equation for the Wigner function is derived and the physical interpretations of the Hartree potential and scattering potential are presented. 
In Sec.~\ref{sec:averageKinetic}, the coarse-grained Wigner function is introduced to derive the quantum kinetic equation in a closed-form. In Sec.~\ref{sec:RedBTE}, the BTE applicable to the cases under discrete impurities is derived. 
Finally, some conclusions are drawn in Sec.~\ref{sec:concl}.

\section{\label{sec:Wigner}Wigner Function and Impurity Potential}

The Wigner function is defined by the Wyle transform of the density operator~\cite{Carruthers1983,Liboff2003book} and expressed by
\begin{eqnarray}
f\left( {{\mathbf{p}},{\mathbf{R}},t} \right) = \int {\frac{{{d^3}r}}{{{{\left( {2\pi } \right)}^3}}}{e^{ - i{\mathbf{p}} \cdot {\mathbf{r}}}}\left\langle {{\mathbf{R}} + \frac{{\mathbf{r}}}{2}} \right|\hat \rho \left( t \right)\left| {{\mathbf{R}} - \frac{{\mathbf{r}}}{2}} \right\rangle } ,
\label{eq:Wigner}
\end{eqnarray}
where  ${\hat \rho \left( t \right)}$ is the density operator of electrons in a semiconductor substrate. 
In the present study, we assume that electrons are so dilute that these electrons are treated as completely independent of the others and the system is dealt with as a single-electron problem.
Hence, the electron density operator is represented by a mixed state such that 
\begin{eqnarray}
\hat \rho \left( t \right) = \frac{1}{\nu }\sum\limits_\alpha  {\left| {{\psi _\alpha }\left( t \right)} \right\rangle \left\langle {{\psi _\alpha }\left( t \right)} \right|} , 
\end{eqnarray}
where $\nu $ is the number of electrons, ${\left| {{\psi _\alpha}\left( t \right)} \right\rangle }$ is the single-electron state-vector and the sum is taken over all electrons~\cite{Tolman1979}.
$\hat \rho \left( t \right) $ satisfies the quantum Liouville equation
\begin{eqnarray}
i \frac{{\partial \hat \rho \left( t \right)}}{{\partial t}} = \left[ {\hat H,\hat \rho \left( t \right)} \right] ,
\label{eq:Liouville}
\end{eqnarray}
where $\hat H$ is the single-electron Hamiltonian and given by 
$\hat H = {{\hat p}^2}/\left( {2m} \right) -e \hat V$, 
where $\hat V$ is the external potential (operator) including the impurity potential, which is self-consistently coupled with the Poisson equation. 
In the present paper, units are chosen such that $\hbar \equiv 1$.
The quantum kinetic equation for the Wigner function is then derived as 
%
\begin{eqnarray}
&&~ \left[ {\frac{\partial }{{\partial t}} + \frac{{\mathbf{p}}}{m} \cdot \frac{\partial }{{\partial {\mathbf{R}}}}} \right]f\left( {{\mathbf{p}},{\mathbf{R}},t} \right) =  - \frac{e}{i}\int {\frac{{{d^3}r{d^3}p'}}{{{{\left( {2\pi } \right)}^3}}}{e^{i\left( {{\mathbf{p'}} - {\mathbf{p}}} \right) \cdot {\mathbf{r}}}}} 
 \nonumber \\
&& \times \left[ {V\left( {{\mathbf{R}} + \frac{{\mathbf{r}}}{2},t} \right) - V\left( {{\mathbf{R}} - \frac{{\mathbf{r}}}{2},t} \right)} \right]f\left( {{\mathbf{p'}},{\mathbf{R}},t} \right).
\label{eq:kinetic}
\end{eqnarray}
%

The discrete impurity density is approximated by a set of delta-functions and expressed by  
\begin{eqnarray}
N_d^ + \left( {\mathbf{R}} \right) = \sum\limits_{i = 1}^{{N_I}} {\delta \left( {{\mathbf{R}} - {{\mathbf{R}}_i}} \right)}  ,
\label{eq:Nd+}
\end{eqnarray}
where $\mathbf{R}_i$ is the position of the $i$-th impurity and 
$N_I$ is the total number of impurities in the substrate. 
The external potential $V\left( {\mathbf{R}} ,t \right)$ in Eq.~\eqref{eq:kinetic} 
satisfies Poisson's equation which is, for later discussions, written by
\begin{eqnarray}
{\nabla ^2}V &=&  - \frac{e}{{{\varepsilon _s}}}\left[ {\left\{ {N_d^ + \left( {\mathbf{R}} \right) - N_{d,long}^ + \left( {\mathbf{R}} \right)} \right\}} \right.  \nonumber \\
&& + \left. \left\{  { N_{d,long}^ + \left( {\mathbf{R}} \right) - n\left( {\mathbf{R}} ,t \right)} \right\} \right],
\label{eq:Poissond}
\end{eqnarray}
where ${N_{d,long}^ + \left( {\mathbf{R}} \right)}$ inserted intentionally on the right-hand side represents the impurity density associated with the long-range part of the impurity Coulomb potential (determined below). $n\left( {{\mathbf{R}},t} \right)$ is
the electron density calculated from the Wigner function by 
\begin{eqnarray}
n\left( {{\mathbf{R}},t} \right) = \int {{d^3}{\mathbf{p}}\,f\left( {{\mathbf{p}},{\mathbf{R}},t} \right)}.
\label{eq:eldW}
\end{eqnarray}
By a proper choice of ${N_{d,long}^ + \left( {\mathbf{R}} \right)}$, 
Eq.~\eqref{eq:Poissond} can be split into the following two equations;
\begin{eqnarray}
{\nabla ^2}{V_{long}} =  - \frac{e}{{{\varepsilon _s}}}\left[ {N_{d,long}^ + \left( {\mathbf{R}} \right) - n\left( {\mathbf{R}},t \right)} \right]
\label{eq:Vlong}
\end{eqnarray}
and 
\begin{eqnarray}
{\nabla ^2}{V_{short}} =  - \frac{e}{{{\varepsilon _s}}}\left[ {N_d^ + \left( {\mathbf{R}} \right) - N_{d,long}^ + \left( {\mathbf{R}} \right)} \right],
\label{eq:Vshort}
\end{eqnarray}
where ${V_{long}}$ and ${V_{short}}$ are the long- and short-range parts of the external potential, respectively, and fulfill the relation $V = V_{long}+V_{short}$.
We would like to point out that $n\left( {{\mathbf{R}},t} \right)$ calculated from Eq.~\eqref{eq:eldW} is regarded as a long-range quantity since $f\left( {{\mathbf{p}},{\mathbf{R}},t} \right)$ is a continuous and smooth function of both $\mathbf{p}$ and $\mathbf{R}$. 

The separation of impurity potential into the long- and short-range parts becomes unique when one considers the system in equilibrium under bulk structures where the charge neutrality condition holds~\cite{Sano2018,Sano2020}. Then, 
$V_{long}$ reduces to a flat potential 
because the long-range part of the impurity potential is completely screened by free electrons and an unscreened part of the potential is excluded from $V_{long}$, owing to the explicit separation of the potential. 
Hence, the right-hand side of Eq.~\eqref{eq:Vlong} must vanish in equilibrium and 
we obtain 
\begin{eqnarray}
N_{d,long}^ + \left( {\mathbf{R}} \right) \approx \sum\limits_{i = 1}^{{N_I}} {n_i^{eq}\left( {\mathbf{R}- \mathbf{R}_i} \right)} .
\label{eq:Nd_long}
\end{eqnarray}
Notice that Eq.~\eqref{eq:Vshort} is now decoupled from the quantum kinetic equation \eqref{eq:kinetic} and closed by itself. This implies that potential modulations induced by screening 
are self-consistently taken into account by Eqs.~\eqref{eq:kinetic} and \eqref{eq:Vlong}. In other words, static screening and dynamical screening are separated under the present framework and the latter is treated as collective motion through the quantum kinetic equation, Eq.~\eqref{eq:kinetic}. 
Assuming that the average impurity density is not extremely large, $N_{d,long}^ +$ is given by a sum of the electron density ${n_i^{eq}}\left( {\mathbf{R}- \mathbf{R}_i} \right)$ modulated around each impurity. Then, we may write ${V_{short}}\left( {\mathbf{R}} \right) = \sum\nolimits_{i = 1}^{{N_I}} {{v_{s}}\left( {\mathbf{R}-\mathbf{R}_i} \right)}$ and   
Eq.~\eqref{eq:Vshort} reduces to Poisson's equation for a {\em single} impurity potential $v_s$, 
\begin{eqnarray}
{\nabla ^2}{v_{s}}=  - \frac{e}{{{\varepsilon _s}}}\left\{ {\delta \left( {{\mathbf{R}- \mathbf{R}_i} } \right) - n_i^{eq}\left( {\mathbf{R}- \mathbf{R}_i} \right)} \right\}. 
\label{eq:Vshort3}
\end{eqnarray}
Clearly, this equation along with appropriate boundary conditions leads to the (static) screened Coulomb potential due to a single impurity at position $\mathbf{R}_i$. When we solve Eq.~\eqref{eq:Vshort3} under the linear approximation for unbounded bulk structure, the usual Yukawa potential is derived~\cite{Mahan2000}.

\section{\label{sec:Kinetic}Quantum Kinetic Equation for Wigner Function}

We now derive the quantum kinetic equation for the Wigner function {\em under discrete impurities}. 
The source term, given by 
the right-hand side of Eq.~\eqref{eq:kinetic}, could be split into two terms in accordance with the long- and short-range parts of the potential 
and we may rewrite Eq.~\eqref{eq:kinetic} as
\begin{eqnarray}
\left[ {\frac{\partial }{{\partial t}} + \frac{{\mathbf{p}}}{m} \cdot \frac{\partial }{{\partial {\mathbf{R}}}}} \right] && f\left( {{\mathbf{p}},{\mathbf{R}},t} \right) 
 \nonumber \\
 = && {S_{long}}\left( {{\mathbf{p}},{\mathbf{R}},t} \right) + {S_{short}}\left( {{\mathbf{p}},{\mathbf{R}},t} \right) ,
\label{eq:kinetic2}
\end{eqnarray}
where ${S_{long}}$ and ${S_{short}}$ represent the source terms associated with ${V_{long}}$ and ${V_{short}}$, respectively.

\subsection{\label{sec:long}Long-range part of the source term}

The long-range part of the source term 
is treated in the usual manner~\cite{Groot1972}. Expanding ${V_{long}}$ in ${S_{long}}\left( {{\mathbf{p}},{\mathbf{R}}}, t \right)$ around $\mathbf{R}$ by the Taylor series, we find 
\begin{eqnarray}
{S_{long}}&&\left( {{\mathbf{p}},{\mathbf{R}},t} \right)  
\nonumber \\
&& ~ = - 2e\sin \left\{ {\frac{1}{2}\left( {\frac{{{\partial ^{(V)}}}}{{\partial {\mathbf{R}}}} \cdot \frac{{{\partial ^{(f)}}}}{{\partial {\mathbf{p}}}} - \frac{{{\partial ^{(V)}}}}{{\partial {\mathbf{p}}}} \cdot \frac{{{\partial ^{(f)}}}}{{\partial {\mathbf{R}}}}} \right)} \right\}
\nonumber \\
&&~~~~~\times {V_{long}}\left( {\mathbf{R}} \right)f\left( {{\mathbf{p}},{\mathbf{R}},t} \right) ,
\label{eq:Slong}
\end{eqnarray}
where $\sin (\cdot)$ is a short-hand notation of its infinite series, and the first and second derivatives inside the argument act on $V_{long}$ and $f$, respectively. 

Notice that this result is exact and no approximation has been made. 
In the present paper, impurity scattering is our main concern so that $V_{long}$ is independent of the electron's momentum. Therefore, the second term inside the argument of the $\sin $ function has essentially no effect. 
However, it is critical to keep this term in Eq.~\eqref{eq:Slong} for being equivalent to the Poisson bracket of Eq.~\eqref{eq:Liouville}, namely,
\begin{eqnarray}
&& {S_{long}}\left( {{\mathbf{p}},{\mathbf{R}},t} \right) \nonumber \\
&& = \int {\frac{{{d^3}r}}{{{{\left( {2\pi } \right)}^3}}}{e^{ - i{\mathbf{p}} \cdot {\mathbf{r}}}}\left\langle {{\mathbf{R}} + \frac{{\mathbf{r}}}{2}} \right|i\left[ {e{{\hat V}_{long}},\hat \rho \left( t \right)} \right]\left| {{\mathbf{R}} - \frac{{\mathbf{r}}}{2}} \right\rangle } . 
\nonumber \\
\end{eqnarray}
Of course, it plays a key role when the Lorentz force associated with the magnetic field is involved. Also, we should mention that the higher-order terms above the lowest-order in Eq.~\eqref{eq:Slong} lead to the renormalization effects of the electronic states in a similar manner to those found in the Kadanoff-Baym equation~\cite{Kadanoff1962,Haug2008}.

\subsection{\label{sec:short}Short-range part of the source term}

Since the length-scale dominant in ${S_{short}}\left( {{\mathbf{p}},{\mathbf{R}},t} \right)$ is in the opposite limit of ${S_{long}}\left( {{\mathbf{p}},{\mathbf{R}},t} \right)$, it is more convenient to work in the Fourier-Laplace space, rather than in the real space.
The Fourier-Laplace-transform of ${S_{short}}$ is expressed by
\begin{eqnarray}
\tilde S&&_{short}\left( {{\mathbf{p}},{\mathbf{k}},\omega } \right) =  - \frac{e}{{i}}\int {\frac{{{d^3}q}}{{{{\left( {2\pi } \right)}^3}}}{{\tilde V}_{short}}\left( {{\mathbf{q}}} \right)} \nonumber \\
&& \times \left[ {\tilde f\left( {{\mathbf{p}} - \frac{{{\mathbf{q}}}}{2},{\mathbf{k}} - {\mathbf{q}},\omega } \right) - \tilde f\left( {{\mathbf{p}} + \frac{{{\mathbf{q}}}}{2},{\mathbf{k}} - {\mathbf{q}},\omega } \right)} \right],  
\nonumber  \\
\label{eq:Sshort}
\end{eqnarray}
where $\tilde S_{short}\left( {{\mathbf{p}},{\mathbf{k}},\omega } \right)$ and ${{\tilde V}_{short}}\left( {\mathbf{q}} \right)$ are, respectively, defined by 
\begin{eqnarray}
 \tilde S&&_{short}\left( {{\mathbf{p}},{\mathbf{k}},\omega } \right)  \nonumber \\
&&   = \int_0^\infty  {dt} \int {{d^3}R{e^{ - i{\mathbf{k}} \cdot {\mathbf{R}} + i\omega t}}{S_{short}}\left( {{\mathbf{p}},{\mathbf{R}},t} \right)} 
\label{eq:Sqomg}
\end{eqnarray}
and
\begin{eqnarray}
{{\tilde V}_{short}}\left( {\mathbf{q}} \right) = \int {{d^3}R{e^{ - i{\mathbf{q}} \cdot {\mathbf{R}}}}{V_{short}}\left( {\mathbf{R}} \right)} .
\label{eq:Vq}
\end{eqnarray}

Equation~\eqref{eq:kinetic2} is formally solved in the Fourier-Laplace space by using the Green function. 
The Wigner function is then given by 
\begin{eqnarray}
 \tilde f && \left( {{\mathbf{p}},{\mathbf{k}},\omega } \right) = i f\left( {{\mathbf{p}},{\mathbf{k}},t = 0} \right) + {G^ + }\left( {{\mathbf{p}},{\mathbf{k}},\omega } \right)  \nonumber \\
&& ~~~~~\times i \left[ {{\tilde S_{long}}\left( {{\mathbf{p}},{\mathbf{k}},\omega } \right) + {\tilde S_{short}}\left( {{\mathbf{p}},{\mathbf{k}},\omega } \right)} \right] 
\label{eq:fexact}
\end{eqnarray}
where ${G^ + }\left( {{\mathbf{p}},{\mathbf{k}},\omega } \right)$ is the retarded Green function and defined by 
\begin{eqnarray}
{G^ + }\left( {{\mathbf{p}},{\mathbf{k}},\omega } \right) = \frac{1}{{\omega  + i\varepsilon  - {\mathbf{p}} \cdot {\mathbf{k}}/m}} .
\label{eq:G+}
\end{eqnarray}
Here, $\varepsilon$ is a positive infinitesimal. 
Although Eq.~\eqref{eq:fexact} is an exact solution, we need to make an approximation to evaluate $\tilde S_{short}$ given by Eq.~\eqref{eq:Sshort}. 
First, we ignore the effects induced by $V_{long}$ during the transition due to impurity scattering, namely,  the intra-collisional field effect (ICFE). 
Hence, the Wigner function in Eq.~\eqref{eq:Sshort} is approximated by 
\begin{eqnarray}
\tilde f\left( {{\mathbf{p}},{\mathbf{k}},\omega } \right) \approx  {G^ + }\left( {{\mathbf{p}},{\mathbf{k}},\omega } \right)  
i{\tilde S_{short}}\left( {{\mathbf{p}},{\mathbf{k}},\omega } \right).
\label{eq:fcoll}
\end{eqnarray}
This approximation may be justified because the collision duration is expected to be short, thanks to the short-range nature of $V_{short}$, so that the retardation effect during scattering should be insignificant unless an external electric field is extremely large. 
In addition, we ignore a transient correlation resulting from the initial condition of the Wigner function $f\left( {{\mathbf{p}},{\mathbf{k}},t=0 } \right)$, because the transport properties driven by the source term of Eq.~\eqref{eq:kinetic2} are our main concerns. 

Substitution of Eq.~\eqref{eq:fcoll} into Eq.~\eqref{eq:Sshort} leads to the following expression for $\tilde S_{short}\left( {{\mathbf{p}},{\mathbf{k}},\omega } \right) $,
\begin{widetext}
\begin{eqnarray}
{{\tilde S}_{short}}\left( {{\mathbf{p}},{\mathbf{k}},\omega } \right) = - e\int {\frac{{{d^3}q}}{{{{\left( {2\pi } \right)}^3}}}{{\tilde V}_{short}}\left( {\mathbf{q}} \right)} 
&&  \left[ {{G^ + }\left( {{\mathbf{p}} - \frac{{\mathbf{q}}}{2},{\mathbf{k}} - {\mathbf{q}},\omega } \right){{\tilde S}_{short}}\left( {{\mathbf{p}} - \frac{{\mathbf{q}}}{2},{\mathbf{k}} - {\mathbf{q}},\omega } \right)} \right.
\nonumber  \\  && 
~~~ - \left. {{G^ + }\left( {{\mathbf{p}} + \frac{{\mathbf{q}}}{2},{\mathbf{k}} - {\mathbf{q}},\omega } \right){{\tilde S}_{short}}\left( {{\mathbf{p}} + \frac{{\mathbf{q}}}{2},{\mathbf{k}} - {\mathbf{q}},\omega } \right)} \right] .
\label{eq:Sqomg1}
\end{eqnarray}
\end{widetext}
This formula plays a similar role to the Lippmann-Schwinger equation in the scattering theory~\cite{RGNewton,JCTaylor,Carruthers1983}. Hence, all higher-order effects associated with short-range  interaction between electrons and impurities such as multiple scattering are included. 
Keeping only the second-order in $\tilde V_{{short}}$, which is equivalent to the Born approximation, ${\tilde S_{short}}\left( {{\mathbf{p}},{\mathbf{k}},\omega } \right)$ is approximated by
\begin{widetext}
\begin{eqnarray}
{{\tilde S}_{short}}\left( {{\mathbf{p}},{\mathbf{k}},\omega } \right) && \approx \frac{{{e^2}}}{i}\int {\frac{{{d^3}q{d^3}q'}}{{{{\left( {2\pi } \right)}^6}}}{{\tilde V}_{short}}\left( { - {\mathbf{q}}} \right){{\tilde V}_{short}}\left( {{\mathbf{q'}}} \right)} 
\nonumber  \\  
&&~~ \times \left\{ {{G^ + }\left( {{\mathbf{p}} + \frac{{\mathbf{q}}}{2},{\mathbf{k}} + {\mathbf{q}},\omega } \right)\left[ {\tilde f\left( {{\mathbf{p}} + \frac{{{\mathbf{q}} - {\mathbf{q'}}}}{2},{\mathbf{k}} + {\mathbf{q}} - {\mathbf{q'}},\omega } \right) - \tilde f\left( {{\mathbf{p}} + \frac{{{\mathbf{q}} + {\mathbf{q'}}}}{2},{\mathbf{k}} + {\mathbf{q}} - {\mathbf{q'}},\omega } \right)} \right]} \right.
\nonumber  \\  
&&~~ - \left. {{G^ + }\left( {{\mathbf{p}} - \frac{{\mathbf{q}}}{2},{\mathbf{k}} + {\mathbf{q}},\omega } \right)\left[ {\tilde f\left( {{\mathbf{p}} - \frac{{{\mathbf{q}} + {\mathbf{q'}}}}{2},{\mathbf{k}} + {\mathbf{q}} - {\mathbf{q'}},\omega } \right) - \tilde f\left( {{\mathbf{p}} - \frac{{{\mathbf{q}} - {\mathbf{q'}}}}{2},{\mathbf{k}} + {\mathbf{q}} - {\mathbf{q'}},\omega } \right)} \right]} \right\} .
\label{eq:Sqomg2}
\end{eqnarray}
\end{widetext}
%

\subsection{\label{sec:QKE}Quantum kinetic equation}
The quantum kinetic equation for the Wigner function is now given by 
\begin{widetext}
\begin{eqnarray}
\left[ { - i\omega  + \frac{{\mathbf{p}}}{m} \cdot \frac{\partial }{{\partial {\mathbf{R}}}} + 2e\sin \left\{ {\frac{1}{2}\left( {\frac{{{\partial ^{(V)}}}}{{\partial {\mathbf{R}}}} \cdot \frac{{{\partial ^{(f)}}}}{{\partial {\mathbf{p}}}} - \frac{{{\partial ^{(V)}}}}{{\partial {\mathbf{p}}}} \cdot \frac{{{\partial ^{(f)}}}}{{\partial {\mathbf{R}}}}} \right)} \right\}{V_{long}}\left( {\mathbf{R}} \right)} \right]f\left( {{\mathbf{p}},{\mathbf{R}},\omega } \right) = {S_{short}}\left( {{\mathbf{p}},{\mathbf{R}},\omega } \right),
\label{eq:kineticF}
\end{eqnarray}
\end{widetext}
where ${S_{short}}\left( {{\mathbf{p}},{\mathbf{R}},\omega } \right)$ is the inverse-Fourier transform of ${{\tilde S}_{short}}\left( {{\mathbf{p}},{\mathbf{k}},\omega } \right)$ expressed by Eq.~\eqref{eq:Sqomg2}.

We would like to stress, however, that Eq.~\eqref{eq:kineticF} is incomplete in the respect that the Fourier transform of $f\left( {{\mathbf{p}},{\mathbf{R}},\omega } \right)$ on the left-hand side of Eq.~\eqref{eq:kineticF} is not consistent with $\tilde f \left( {{\mathbf{p}},{\mathbf{k}},\omega } \right)$ in ${{ S}_{short}} \left( {{\mathbf{p}},{\mathbf{R}},\omega } \right) $: The ICFE associated with the (long-range) spatial variation of $V_{long}\left( {\mathbf{R}} \right)$ is ignored in $\tilde f \left( {{\mathbf{p}},{\mathbf{k}},\omega } \right)$ of ${{ S}_{short}} \left( {{\mathbf{p}},{\mathbf{R}},\omega } \right) $ because of the approximation employed by Eq.~\eqref{eq:fcoll}. 
Therefore, the following two important facts must be considered. 
First, $\tilde f \left( {{\mathbf{p}},{\mathbf{k}},\omega } \right)$ in Eq.~\eqref{eq:Sqomg2} is regarded as the Fourier-Laplace transform of $f\left( {{\mathbf{p}},{\mathbf{R}},t } \right)$ as if the electrostatic potential were fixed with the ``local'' value $V_{long}\left( {\mathbf{R}} \right)$ during collision duration. 
Second, the Markov approximation is inevitable at this stage because $\tilde f \left( {{\mathbf{p}},{\mathbf{k}},\omega } \right)$ of ${{ S}_{short}} \left( {{\mathbf{p}},{\mathbf{R}},\omega } \right) $ is not updated during collision duration consistently with $V_{long}\left( {\mathbf{R}} \right)$. 
As a result, $\omega $-dependence in the retarded Green function ${G^ + }\left( {{\mathbf{p}},{\mathbf{k}},\omega } \right)$ is decoupled from that of $\tilde f \left( {{\mathbf{p}},{\mathbf{k}},\omega } \right)$ in Eq.~\eqref{eq:Sqomg2}.
In other words, the quantum kinetic equation~\eqref{eq:kineticF} is not actually closed for the Wigner function and we need to complement another relation between $f\left( {{\mathbf{p}},{\mathbf{R}},\omega } \right)$ of the left-hand side of Eq.~\eqref{eq:kineticF} and $\tilde f \left( {{\mathbf{p}},{\mathbf{k}},\omega } \right)$ in ${S_{short}}\left( {{\mathbf{p}},{\mathbf{R}},\omega } \right)$ of the right-hand side.

\section{\label{sec:averageKinetic}Kinetic Equation for Coarse-grained Wigner Function}
\subsection{\label{sec:aveWigner}Coarse-grained Wigner function}
To close the quantum kinetic equation \eqref{eq:kineticF} for the Wigner function, we introduce the spatial average (coarse-graining) of the Wigner function with respect to the weight function $g\left( {\mathbf{R}} \right)$, which is normalized to unity over the space. 
In fact, coarse-graining has been often introduced in past studies to explain the appearance of irreversibility in the transport equations, although in most cases it has been applied to the time or energy domain~~\cite{Jancel2013book,Kreuzer1981book}. In the present study, however, we define the coarse-grained Wigner function in space by
\begin{eqnarray}
\left\langle {f\left( {{\mathbf{p}},{\mathbf{R}},\omega } \right)} \right\rangle  = \int_{} {{d^3}R' g\left( {{\mathbf{R'}}} \right)f\left( {{\mathbf{p}},{\mathbf{R}} - {\mathbf{R'}},\omega } \right)} ,
\label{eq:Wignerw}
\end{eqnarray}
and $g{\left( {{\mathbf{R}}} \right)}$ is assumed to have the following expression
\begin{eqnarray}
g\left( {\mathbf{R}} \right) = \frac{{{q_c}^3}}{{{{\left( {2\pi } \right)}^{3/2}}}}{e^{ - \frac{{{q_c}^2{{\mathbf{R}}^2}}}{2}}}.
\label{eq:weight}
\end{eqnarray}
Here, $q_{c}$ is the inverse of the screening length determined by Poisson's equation for the short-range part of the potential, as given by Eq.~\eqref{eq:Vshort3}. Since $q_{c}$ is dependent on the modulated electron density $n_i^{eq}\left( {\mathbf{R}- \mathbf{R}_i} \right)$ around the impurity at $\mathbf{R}_i$, it is also dependent on $\mathbf{R}$ through $V_{long}\left( {\mathbf{R}} \right)$. 
Notice that Eq.~\eqref{eq:Wignerw} is also expressed by
\begin{eqnarray}
\left\langle {f\left( {{\mathbf{p}},{\mathbf{R}},\omega } \right)} \right\rangle  &=& \int {\frac{{{d^3}k}}{{{{\left( {2\pi } \right)}^3}}}{e^{i{\mathbf{k}} \cdot {\mathbf{R}}}}\tilde g\left( {\mathbf{k}} \right)\tilde f\left( {{\mathbf{p}},{\mathbf{k}},\omega } \right)}  
 \nonumber \\
&\approx & \frac{1}{{\Delta {\Omega _{\mathbf{R}}}}}\tilde f\left( {{\mathbf{p}},{\mathbf{k}} = 0,\omega } \right) ,
\label{eq:aveW}
\end{eqnarray}
where in the last line we used the fact that $\tilde g\left( {\mathbf{k}} \right)$ is given by 
%
$\tilde g\left( {\mathbf{k}} \right) = \exp \left( { - {{\mathbf{k}}^2}/2{q_c}^2} \right)$
%
and sharply peaked at $\mathbf{k} =0$. $\Delta {\Omega _{\mathbf{R}}}$ is the volume with the screening length $q_c^{-1}$ and given by $\Delta {\Omega _{\mathbf{R}}} = {\left( {2\pi } \right)^{3/2}}/{q_c}^3$ and, thus, 
$\Delta {\Omega _{\mathbf{R}}}$ also becomes dependent on position $\mathbf{R}$ through $q_c^{-1}$. It is clear from Eq.~\eqref{eq:aveW} that $\left\langle {f\left( {{\mathbf{p}},{\mathbf{R}},\omega } \right)} \right\rangle $ is indeed the Wigner function averaged over the (macroscopically small) volume at $\mathbf{R}$. As a result, $\tilde f\left( {{\mathbf{p}},{\mathbf{k}} = 0,\omega } \right)$ is diagonal in momentum space and $\left\langle {f\left( {{\mathbf{p}},{\mathbf{R}},\omega } \right)} \right\rangle$ becomes positive semi-definite. 
Furthermore, as seen from Eq.~\eqref{eq:Wignerw}, $\tilde f\left( {{\mathbf{p}},{\mathbf{k}} = 0,\omega } \right)$ is the Fourier-transform of the Wigner function centered at the position $\mathbf{R}$. Hence, $\tilde f\left( {{\mathbf{p}},{\mathbf{k}} = 0,\omega } \right)$ may be regarded as the Fourier-transformed Wigner function that is obtained under the assumption as if the electrostatic potential were fixed with the value of $V_{long}\left( {\mathbf{R}} \right)$ over the volume $\Omega$.  
This is exactly the same interpretation employed for the Fourier-transformed Wigner function in ${S_{short}}\left( {{\mathbf{p}},{\mathbf{R}},\omega } \right)$ of Eq.~\eqref{eq:kineticF}. Therefore, 
Eq.~\eqref{eq:aveW} plays a complementary relation to close Eq.~\eqref{eq:kineticF}.

\subsection{\label{sec:aveKinetic}Quantum kinetic equation in a closed-form}
We now impose the spatial average on both sides of Eq.~\eqref{eq:kineticF}. Noting that the operation of space differentiation commutes with the averaging operation, we immediately obtain
%
\begin{eqnarray}
&&
\left( {- i \omega  + \frac{{\mathbf{p}}}{m} \cdot \frac{\partial }{{\partial {\mathbf{R}}}}} \right)\left\langle {f\left( {{\mathbf{p}},{\mathbf{R}},\omega } \right)} \right\rangle  
\nonumber \\   &&~~~~~~ 
 + 2e\sin \left\{ {\frac{1}{2}\left( {\frac{{{\partial ^{(V)}}}}{{\partial {\mathbf{R}}}} \cdot \frac{{{\partial ^{(f)}}}}{{\partial {\mathbf{p}}}} - \frac{{{\partial ^{(V)}}}}{{\partial {\mathbf{p}}}} \cdot \frac{{{\partial ^{(f)}}}}{{\partial {\mathbf{R}}}}} \right)} \right\}
\nonumber \\   && \times 
\left\langle {{V_{long}}\left( {\mathbf{R}} \right)f\left( {{\mathbf{p}},{\mathbf{R}},\omega } \right)} \right\rangle  = \left\langle {{S_{short}}\left( {{\mathbf{p}},{\mathbf{R}},\omega } \right)} \right\rangle .
\label{eq:aveQKT}
\end{eqnarray}
%
Notice that $V_{long}\left( {\mathbf{R}} \right)$ varies in space with the length-scale greater than $q_{c}^{-1}$, whereas the space average is taken over the region smaller than $q_{c}^{-1}$. Hence, the space-averaging of $V_{long}\left( {\mathbf{R}} \right)$ has little effect and we can approximate 
\begin{eqnarray}
\left\langle {{V_{long}}\left( {\mathbf{R}} \right)f\left( {{\mathbf{p}},{\mathbf{R}},\omega } \right)} \right\rangle  \approx {V_{long}}\left( {\mathbf{R}} \right)\left\langle {f\left( {{\mathbf{p}},{\mathbf{R}},\omega } \right)} \right\rangle .
\label{eq:split}
\end{eqnarray}

On the other hand, the right-hand side of Eq.~\eqref{eq:aveQKT} becomes
\begin{widetext}
\begin{eqnarray}
&& \left\langle {{S_{short}}\left( {{\mathbf{p}},{\mathbf{R}},\omega } \right)} \right\rangle  = \frac{{{e^2}}}{i}\frac{1}{{{\Omega ^2}}}\sum\limits_{{\mathbf{q}},{\mathbf{q'}}} {{{\tilde V}_{short}}\left( { - {\mathbf{q}}} \right){{\tilde V}_{short}}\left( {{\mathbf{q'}}} \right)} \frac{1}{{\Delta {\Omega _{\mathbf{R}}}}}\left\{ {{G^ + }\left( {{\mathbf{p}} + \frac{{\mathbf{q}}}{2},{\mathbf{q}},\omega } \right)} \right.\left[ {\tilde f\left( {{\mathbf{p}} + \frac{{{\mathbf{q}} - {\mathbf{q'}}}}{2},{\mathbf{q}} - {\mathbf{q'}},\omega } \right)} \right.
  \nonumber \\
&&~ \left. { - \tilde f\left( {{\mathbf{p}} + \frac{{{\mathbf{q}} + {\mathbf{q'}}}}{2},{\mathbf{q}} - {\mathbf{q'}},\omega } \right)} \right] - \left. {{G^ + }\left( {{\mathbf{p}} - \frac{{\mathbf{q}}}{2},{\mathbf{q}},\omega } \right)\left[ {\tilde f\left( {{\mathbf{p}} - \frac{{{\mathbf{q}} + {\mathbf{q'}}}}{2},{\mathbf{q}} - {\mathbf{q'}},\omega } \right) - \tilde f\left( {{\mathbf{p}} - \frac{{{\mathbf{q}} - {\mathbf{q'}}}}{2},{\mathbf{q}} - {\mathbf{q'}},\omega } \right)} \right]} \right\},
 \nonumber \\
\label{eq:Sqomg4}
\end{eqnarray}
\end{widetext}
where $\Omega $ is the volume of the substrate and integral over the momentum is now replaced by the sum by following the box normalization rule. Also, the following relationship is used; 
\begin{eqnarray}
\left\langle {{e^{i{\mathbf{k}} \cdot {\mathbf{R}}}}} \right\rangle  = {e^{i{\mathbf{k}} \cdot {\mathbf{R}}}}\tilde g\left( {\mathbf{k}} \right) .
\end{eqnarray}
%
%
Because of Eq.~\eqref{eq:aveW}, $\left\langle {f\left( {{\mathbf{p}},{\mathbf{R}},\omega } \right)} \right\rangle$ on the left-hand side of Eq.~\eqref{eq:aveQKT} is diagonal in momentum space and, thus, 
Eq.~\eqref{eq:Sqomg4} is singular with respect to the diagonal components of $\tilde f \left( {{\mathbf{p}},{\mathbf{k}},\omega } \right)$~\cite{vanHove1957}. 
Hence, we may write  
\begin{eqnarray}
\tilde f\left( {{\mathbf{p}} \pm \frac{{{\mathbf{q}} \pm {\mathbf{q'}}}}{2},{\mathbf{q}} - {\mathbf{q'}},\omega } \right) \approx {\delta _{{\mathbf{q}},\;{\mathbf{q'}}}}\tilde f\left( {{\mathbf{p}} \pm \frac{{{\mathbf{q}} \pm {\mathbf{q}}}}{2},0,\omega } \right).  \nonumber \\
\label{eq:diagW}
\end{eqnarray}
Notice that $\omega$-dependence of $G^{+}\left( {{\mathbf{p}},{\mathbf{q}},\omega } \right)$ represents the energy change during collision duration due to impurity scattering, whereas that of $\tilde f\left( {{\mathbf{p}},{\mathbf{k}},\omega } \right)$ represents the energy change during electron's drift motion induced by the left-hand side of Eq.~\eqref{eq:aveQKT}. Therefore, $\omega$-dependences of $G^{+}\left( {{\mathbf{p}},{\mathbf{q}},\omega } \right)$ and $\tilde f\left( {{\mathbf{p}},{\mathbf{k}},\omega } \right)$ are not correlated with each other, as discussed in \ref{sec:QKE}. 
Hence, we may set $\omega = 0$ in $G^{+}\left( {{\mathbf{p}},{\mathbf{q}},\omega } \right)$ because impurity scattering is static and we obtain 
\begin{eqnarray}
&& {G^ + }\left( {{\mathbf{p}} \pm \frac{{\mathbf{q}}}{2},{\mathbf{q}},\omega  = 0} \right) = \frac{1}{{i\varepsilon  \mp \left( {{E_{{\mathbf{p}} \pm {\mathbf{q}}}} - {E_{\mathbf{p}}}} \right)}}
 \nonumber \\
&&  
~~\xrightarrow{{\varepsilon  \to {0^ + }}} \mp P\frac{1}{{{E_{{\mathbf{p}} \pm {\mathbf{q}}}} - {E_{\mathbf{p}}}}} - i\pi \delta \left( {{E_{{\mathbf{p}} \pm {\mathbf{q}}}} - {E_{\mathbf{p}}}} \right) ,
\label{eq:G+2}
\end{eqnarray}
where ${E_{\mathbf{p}}}$ is electron's energy with momentum $\mathbf{p}$. 
Consequently, the coarse-grained source term $\left\langle {{S_{short}}\left( {{\mathbf{p}},{\mathbf{R}},\omega } \right)} \right\rangle $ is given by
\begin{eqnarray}
&& \left\langle {{S_{short}}\left( {{\mathbf{p}},{\mathbf{R}},\omega } \right)} \right\rangle  = 2\pi \frac{e^{2}}{{{\Omega ^2}}}\sum\limits_{\mathbf{q}} {{{\left| {{{\tilde V}_{short}}\left( {\mathbf{q}} \right)} \right|}^2}\delta \left( {{E_{{\mathbf{p}} \pm {\mathbf{q}}}} - {E_{\mathbf{p}}}} \right)} 
 \nonumber \\
&&  ~~~~~\times \frac{1}{{\Delta {\Omega _{\mathbf{R}}}}}\left[ {\tilde f\left( {{\mathbf{p}} + {\mathbf{q}},0,\omega } \right) - \tilde f\left( {{\mathbf{p}},0,\omega } \right)} \right] .
\label{eq:G+2p}
\end{eqnarray}
Using Eq.~\eqref{eq:aveW}, the quantum kinetic equation~\eqref{eq:kineticF} is now closed for the coarse-grained Wigner function ${\left\langle {f\left( {{\mathbf{p}},R,\omega } \right)} \right\rangle }$ and we obtain  
\begin{widetext}
\begin{eqnarray}
&& \left[ { - i\omega  + \frac{{\mathbf{p}}}{m} \cdot \frac{\partial }{{\partial {\mathbf{R}}}} + 2e\sin \left\{ {\frac{1}{2}\left( {\frac{{{\partial ^{(V)}}}}{{\partial {\mathbf{R}}}} \cdot \frac{{{\partial ^{(f)}}}}{{\partial {\mathbf{p}}}} - \frac{{{\partial ^{(V)}}}}{{\partial {\mathbf{p}}}} \cdot \frac{{{\partial ^{(f)}}}}{{\partial {\mathbf{R}}}}} \right)} \right\}{V_{long}}\left( {\mathbf{R}} \right)} \right]\left\langle {f\left( {{\mathbf{p}},\mathbf{R},\omega } \right)} \right\rangle  
\nonumber \\
&&~~~~~~~~~~~~~~~~~~~~~~~~~~~~~~~~~~~~~~~~~~~~~~~~~~~~~~
= \sum\limits_{\mathbf{q}} {{W_{imp}}\left( {{\mathbf{p}},{\mathbf{p}} + {\mathbf{q}}} \right)} \left[ {\left\langle {f\left( {{\mathbf{p}} + {\mathbf{q}},\mathbf{R},\omega } \right)} \right\rangle  - \left\langle {f\left( {{\mathbf{p}},\mathbf{R},\omega } \right)} \right\rangle } \right],
\label{eq:QKTfinal}
\end{eqnarray}
\end{widetext}
where the transition probability ${W_{imp}}\left( {{\mathbf{p}},{\mathbf{p}} + {\mathbf{q}}} \right)$ between electron's momentum $\mathbf{p}$ and ${\mathbf{p}} + {\mathbf{q}}$ via short-range impurity scattering is defined by 
\begin{eqnarray}
{W_{imp}}\left( {{\mathbf{p}},{\mathbf{p}} + {\mathbf{q}}} \right) = 2\pi \frac{1}{{{\Omega ^2}}}{\left| {e{{\tilde V}_{short}}\left( {\mathbf{q}} \right)} \right|^2}\delta \left( {{E_{{\mathbf{p}} + {\mathbf{q}}}} - {E_{\mathbf{p}}}} \right).
\nonumber \\
\label{eq:Wimp}
\end{eqnarray}
The square of the scattering potential is expressed by
\begin{eqnarray}
&& \frac{1}{{{\Omega ^2}}}{\left| {e{{\tilde V}_{short}}\left( {\mathbf{q}} \right)} \right|^2} 
\nonumber \\
&&~~~ = \frac{1}{{{\Omega ^2}}}{\left| {e{{\tilde v}_s}\left( {\mathbf{q}} \right)} \right|^2}\sum\limits_{i = 1}^{{N_I}} {\left[ {1 + \sum\limits_{j \ne i}^{{N_I}} {{e^{ - i{\mathbf{q}} \cdot \left( {{{\mathbf{R}}_i} - {{\mathbf{R}}_j}} \right)}}} } \right]}   
\nonumber \\
&&~~~\approx  \frac{{{{\bar n}_{imp}}}}{\Omega }{\left| {e{{\tilde v}_s}\left( {\mathbf{q}} \right)} \right|^2} ,
\label{eq:impVs} 
\end{eqnarray}
where we used the fact that the zero-Fourier component of $\tilde v_s \left( {\mathbf{q}} \right)$ 
is already excluded so that the phase interference among multiple impurities represented by the second term inside the bracket of the second line is negligible. ${{\bar n}_{imp}}$ is the average impurity density and defined by ${{\bar n}_{imp}} = {N_I}/\Omega $.
Then, Eq.~\eqref{eq:impVs} exactly coincides with the expression used for impurity scattering under the conventional Wigner function framework. 

This expression is, however, inappropriate under discrete impurities. 
Noting that 
\begin{eqnarray}
\frac{1}{\Omega }e{{\tilde V}_{short}}\left( {\mathbf{q}} \right) &=& \frac{1}{\Omega }\int_{\Omega } {{d^3}R\,{e^{ - i{\mathbf{q}} \cdot {\mathbf{R}}}}e{V_{short}}\left( {\mathbf{R}} \right)}  
  \nonumber \\
&=& \left\langle {{\mathbf{p}} + {\mathbf{q}}} \right| e{{\hat V}_{short}}\left| {\mathbf{p}} \right\rangle ,
\label{eq:scatkernel}
\end{eqnarray}
the transition matrix associated with the short-range impurity potential is evaluated by electron's plane wave 
$\left\langle {\mathbf{R}} \right.\left| {\mathbf{p}} \right\rangle  = {e^{i{\mathbf{R}} \cdot {\mathbf{p}}}}/\sqrt \Omega  $, 
which spreads over the entire substrate $\Omega $. 
On the other hand, the transition probability $W_{imp}\left( {{\mathbf{p}},{\mathbf{p}} + {\mathbf{q}}} \right) $ is derived from $\left\langle {{S_{short}}\left( {{\mathbf{p}},{\mathbf{R}},\omega } \right)} \right\rangle$ under the assumption that the electrostatic potential is fixed with a ``local'' value $V_{long}\left( {\mathbf{R}} \right)$ during collision duration. Hence, Eq.~\eqref{eq:impVs} should be interpreted only locally in space. Namely, we may regard it as the transition probability obtained for the case as if impurities were distributed over the substrate consistently with the ``local'' long-range potential $V_{long}\left( {\mathbf{R}} \right)$. As a result, $N_{I}$ in Eq.~\eqref{eq:impVs} is replaced by $N_{d,long}^ + \left( {\mathbf{R}} \right) \times \Omega$ and Eq.~\eqref{eq:Wimp} becomes
\begin{eqnarray}
&& {W_{imp}}\left( {{\mathbf{p}},{\mathbf{p}} + {\mathbf{q}} } \right) 
\nonumber \\
&& ~~~~~ = 2\pi N_{d,long}^ + \left( {\mathbf{R}} \right)\frac{1}{\Omega }{\left| {e{{\tilde v}_s}\left( {\mathbf{q}} \right)} \right|^2}\delta \left( {{E_{{\mathbf{p}} + {\mathbf{q}}}} - {E_{\mathbf{p}}}} \right) .
\label{eq:Wimp2}
\end{eqnarray}
%

The quantum kinetic equation derived here is nearly identical to the one widely used for quantum transport simulations for semiconductor nanostructures under the Wigner function scheme~\cite{Ferry2018,Frensley1987a,Nedjalkov2004,Querlioz2006,Sverdlov2005}. 
Yet, there are a few critical differences between the present theory and the conventional Wigner function scheme: One is in the drift term of the left-hand side of Eq.~\eqref{eq:QKTfinal}, in which the whole external potential $V\left( {\mathbf{R}} \right)$ in the conventional scheme is now replaced by $V_{long} {\left( {\mathbf{R}} \right)}$. Hence, as mentioned earlier, the self-consistent Hartree potential is given by $V_{long} {\left( {\mathbf{R}} \right)}$, rather than $V {\left( {\mathbf{R}} \right)}$.
Another point is the fact that the collision integral associated with $V_{short} {\left( {\mathbf{R}} \right)}$ is naturally {\em derived} from the quantum Liouville equation in the present theory. Since the collision integral in the conventional scheme is introduced in an {\em ad hoc} manner along with the {\it a priori} screening potential, the short-range scattering potential is double-counted in both the Hartree potential and the collision integral. 
Furthermore, the Wigner function in the conventional scheme is replaced by the coarse-grained Wigner function. 
Although the Wigner function can be negative where quantum phase interference is significant, the coarse-grained Wigner function in the present theory is positive semi-definite and, thus, it can be interpreted as a probability density.

\section{\label{sec:RedBTE}Reduction to the Boltzmann transport equation}
The reduction of Eq.~\eqref{eq:QKTfinal} to the BTE is straightforward. 
Since ${V_{long}}\left( {\mathbf{R}} \right)$ varies gradually in space, the first term of the series on the left-hand side of Eq.~\eqref{eq:QKTfinal} is most significant. 
Keeping only the lowest-order term, we obtain the drift term which is similar to the expression of the BTE with no magnetic field;
\begin{eqnarray}
&& 2e\sin \left\{ {\frac{1}{2}\left( {\frac{{{\partial ^{(V)}}}}{{\partial {\mathbf{R}}}} \cdot \frac{{{\partial ^{(f)}}}}{{\partial {\mathbf{p}}}} - \frac{{{\partial ^{(V)}}}}{{\partial {\mathbf{p}}}} \cdot \frac{{{\partial ^{(f)}}}}{{\partial {\mathbf{R}}}}} \right)} \right\}
{V_{long}}\left( {\mathbf{R}} \right)
\nonumber \\
&&~ \times \left\langle {f\left( {{\mathbf{p}},R,\omega } \right)} \right\rangle \approx e\frac{{\partial {V_{long}}\left( {\mathbf{R}} \right)}}{{\partial {\mathbf{R}}}} \cdot \frac{\partial }{{\partial {\mathbf{p}}}}\left\langle {f\left( {{\mathbf{p}},R,\omega } \right)} \right\rangle,
\label{eq:drift}
\end{eqnarray}
This simply reflects the fact that quantum phase interference between the incident and reflected waves induced by ${{V_{long}}\left( {\mathbf{R}} \right)}$ is ignored because of gradual variations of the potential. It should be pointed out, however, that this approximation breaks down if the whole potential $V\left( {\mathbf{R}} \right)$, instead of ${{V_{long}}\left( {\mathbf{R}} \right)}$, is used as the Hartree potential. Consequently, Eq.~\eqref{eq:QKTfinal} reduces to the BTE, 
\begin{eqnarray}
&& \left[ { - i\omega  + \frac{{\mathbf{p}}}{m} \cdot \frac{\partial }{{\partial {\mathbf{R}}}} + e\frac{{\partial {V_{long}}\left( {\mathbf{R}} \right)}}{{\partial {\mathbf{R}}}} \cdot \frac{\partial }{{\partial {\mathbf{p}}}}} \right]\left\langle {f\left( {{\mathbf{p}},R,\omega } \right)} \right\rangle  
\nonumber \\
&& = \sum\limits_{\mathbf{q}} {{W_{imp}}\left( {{\mathbf{p}},{\mathbf{p}} + {\mathbf{q}}} \right)} \left[ {\left\langle {f\left( {{\mathbf{p}} + {\mathbf{q}},R,\omega } \right)} \right\rangle  - \left\langle {f\left( {{\mathbf{p}},R,\omega } \right)} \right\rangle } \right] ,
\nonumber \\
\label{eq:exBTE}
\end{eqnarray}

Although the above BTE is almost identical to the conventional BTE with impurity scattering, we should regard it as an extended version of the BTE applicable to discrete impurities. 
Since ${{V_{long}}\left( {\mathbf{R}} \right)}$ is self-consistently coupled with Poisson's equation, the discrete nature of impurities could be represented as potential fluctuations when the screening of impurities by electrons is incomplete. 
Furthermore, the transition probability due to impurities is evaluated by the {\em local} impurity density $N_{d,long}^ + \left( {\mathbf{R}} \right)$, rather than the average impurity density $\bar n_{imp}\left( {\mathbf{R}} \right)$. Therefore, impurity scattering becomes localized around impurities, and the discreteness of impurities is also reflected in the collision integral. 
The above points can be applied to the Monte Carlo simulations based on the BTE and we have recently confirmed that the present scheme is indeed able to simulate potential fluctuations and localized scattering events under discrete impurities such that the same transport properties in bulk structures can be reproduced as those with the BTE under the long-wavelength limit~\cite{Sano2021x}.  
Last, yet most importantly, it should be stressed again that the present derivation is possible only because we do not rely on self-averaging over the random impurity configurations. 

Finally, we would like to mention that a similar treatment can be applied to the long-range part of the Coulomb interaction among electrons. In such cases, the Coulomb interaction induces ``dynamical'' potential fluctuations, namely, plasma oscillations~\cite{Uechi2008,Fukui2008,Nakanishi2009}, and those collective motions are taken into account through the Poisson equation in a self-consistent manner, rather than the dynamical dielectric function of the scattering potential. This is also consistent with the fact that the plasma oscillation among electrons results from the long-range part of the Coulomb potential $V_{long}\left( {\mathbf{R}} \right)$.

\section{\label{sec:concl}Conclusion}
We have derived a quantum kinetic equation for the Wigner function under discrete impurities from the quantum Liouville equation in a closed-form. This has been achieved by explicitly separating the electrostatic Coulomb potential into the long- and short-range parts and taking into account the self-consistent coupling with Poisson's equation. 
We have shown that the collision integral associated with impurity scattering as well as the usual drift term is {\em derived} on an equal footing and, thus, the conventional treatment of impurity scattering under the Wigner function scheme is inconsistent because the collision integral is introduced in an {\it ad hoc} way and  the short-range part of the impurity potential is double-counted. Furthermore, introducing the coarse-grained Wigner function in space, the quantum kinetic equation reduces to a closed-form with the coarse-grained Wigner function and becomes consistent with the approximations implicitly imposed on the kinetic equation. Also, the Wigner function becomes positive semi-definite and, thus, can be interpreted as a probability density in the present framework. 
The BTE has been derived without imposing the long-wavelength limit of the Hartree potential and the self-averaging over random impurity configurations, which have been essential ingredients to {\em derive} the BTE with impurity scattering. The derived BTE is appropriate to take into account the discrete nature of impurities and applicable to the anaysis of electron transport under semiconductor nanostructures.




%

\end{document}